\documentclass{article}

\usepackage{PRIMEarxiv}

\usepackage[utf8]{inputenc} 
\usepackage[T1]{fontenc}    
\usepackage{hyperref}       
\usepackage{url}            
\usepackage{booktabs}       
\usepackage{amsfonts}       
\usepackage{nicefrac}       
\usepackage{microtype}      
\usepackage{lipsum}
\usepackage{fancyhdr}       
\usepackage{graphicx}       
\graphicspath{{media/}}     

\title{HUNIS: High-Performance Unsupervised Nuclei Instance Segmentation
}

\author{
  Vasileios Magoulianitis \\
  Electrical and Computer Engineering \\
  University of Southern California \\
  Los Angeles, CA, USA\\
  \texttt{magoulia@usc.edu} \\
    \And
  Yijing Yang \\
  Electrical and Computer Engineering \\
  University of Southern California \\
  Los Angeles, CA, USA\\
  \texttt{yijingya@usc.edu} \\
    \And
  C.-C. Jay Kuo \\
  Electrical and Computer Engineering \\
  University of Southern California \\
  Los Angeles, CA, USA\\
  \texttt{cckuo@sipi.usc.edu} \\
  }

\begin{document}
\maketitle

\begin{abstract}
A high-performance unsupervised nuclei instance segmentation (HUNIS)
method is proposed in this work. HUNIS consists of two-stage block-wise
operations. The first stage includes: 1) adaptive thresholding of pixel
intensities, 2) incorporation of nuclei size/shape priors and 3) removal
of false positive nuclei instances. Then, HUNIS conducts the second
stage segmentation by receiving guidance from the first one.  The second
stage exploits the segmentation masks obtained in the first stage and
leverages color and shape distributions for a more accurate
segmentation.  The main purpose of the two-stage design is to provide
pixel-wise pseudo-labels from the first to the second stage. This
self-supervision mechanism is novel and effective.  Experimental results
on the MoNuSeg dataset show that HUNIS outperforms all other
unsupervised methods by a substantial margin.  It also has a competitive
standing among state-of-the-art supervised methods. 
\end{abstract}

\section{Introduction}\label{sec:Intro}

Medical imaging is one of the fields that benefit a lot from the
advancement of modern AI algorithms.  They enable the computer aided
diagnosis (CAD) tools to serve as physician's assistants. In particular,
CAD in digital pathology becomes a fast growing area since it is
conducive to cancer diagnosis and assessment. As part of this process,
nuclei segmentation provides important visual cues, such as molecular
morphological information \cite{gurcan2009histopathological} to expert
pathologists. 

Generally speaking, nuclei instance segmentation is an indispensable
task in histology images reading for cancer assessment.  Its automation
is of high significance for pathologists' reading process. Hematoxylin
and Eosin (H\&E) staining has been used for years in histology to reveal
the underlying nuclei structure. Variations along this process,
especially for images coming from different laboratories that use
different protocols and scanners, may affect nuclei color and texture.
Manual segmentation of histology images carried out by expert
pathologists is a labor-intensive and time-consuming task, also subject
to high inter-observer variability \cite{caicedo2019nucleus}. Thus, a
sufficiently large amount of annotated data is in paucity. 

A high-performance unsupervised nuclei instance segmentation (HUNIS)
method is proposed in this work. HUNIS consists of two-stage block-wise
operations, where the first stage provides an initial segmentation
result and yields pixel-wise pseudo-labels for the second stage. This
self-supervision mechanism is novel and effective.  It is shown by
experimental results that HUNIS outperforms other unsupervised methods
by a large margin.  Besides, HUNIS is highly competitive with
state-of-the-art supervised methods.  The rest of the paper is organized
as follows. Related work is reviewed in Sec.  \ref{sec:review}. HUNIS is
presented in Sec.  \ref{sec:PropMeth}.  Experimental results are shown
in Sec.  \ref{sec:Results}.  Concluding remarks are given in Sec.
\ref{sec:Concls}. 

\section{Review of Related Work}\label{sec:review}

Before the advent of the deep learning (DL) paradigm, earlier methods
addressed this segmentation problem with no supervision. Examples
include: Adaptive thresholding \cite{xue2011t,lu2012robust,
magoulianitis2021unsupervised}, clustering \cite{hafiane2008fuzzy},
active contours \cite{al2016white, roula2004evolutionary}, and graph
cuts \cite{xu2021unsupervised}. Another popular method is the watershed
algorithm \cite{shen2015segmenting, veta2011marker}, which is often used
as a post-processing step, where research was mainly focused on finding proper markers to initialize
the segmentation process. 

In recent years, DL solutions are prevalent \cite{kumar2017dataset,
oda2018besnet, graham2019hover, lal2021nucleisegnet, schlemper2019attention}. They attempt to
handle multi-scale appearances of nuclei through separate branches of
the networks and negative effects of hard samples through customized
losses \cite{zhou2019cia, xie2020instance, graham2019hover}.  A
self-supervised learning method was proposed by Sahasrabudhe {\em et
al.} \cite{sahasrabudhe2020self}, that regularizes the encoder model
implicitly with scale. Despite their effectiveness, it is challenging
for DL models to generalize and transfer learned models from training to
testing domains. Given the small-sized publicly available datasets and
nuclei variations across different organs, DL solutions have their
limitations. 

Lately, unsupervised methods have shown promising performance on the
nuclei instance segmentation task, e.g.,
\cite{magoulianitis2021unsupervised, liu2020unsupervised, hsu2021darcnn,
hou2019robust}. Among them, \cite{liu2020unsupervised, hsu2021darcnn,
hou2019robust} are DL-based methods.  \cite{liu2020unsupervised} and \cite{
hsu2021darcnn} adopt domain adaptation and model regularization, while
\cite{hou2019robust} uses generative adversarial networks (GANs) to
synthesize histology images for the nuclei segmentation model. Yet, their
performance is far inferior to that of supervised DL methods. On the other hand, the CBM method \cite{magoulianitis2021unsupervised} is a non-DL solution, offering a transparent pipeline for addressing the problem and requiring no training data.  

In this work, we devise a two-stage unsupervised processing pipeline, namely HUNIS. The first stage consists of a novel adaptive thresholding operation and a false positive (FP) nuclei removal module, to obtain an initial segmentation output. Then, the first stage's output is used to provide pixel-wise pseudo-label to guide the second stage processing for a more accurate segmentation. This self-supervision mechanism is novel and effective as demonstrated by experimental results in Sec.
\ref{sec:Results}.

\begin{figure*}[t]
\begin{center}
\includegraphics[width=0.8\linewidth]{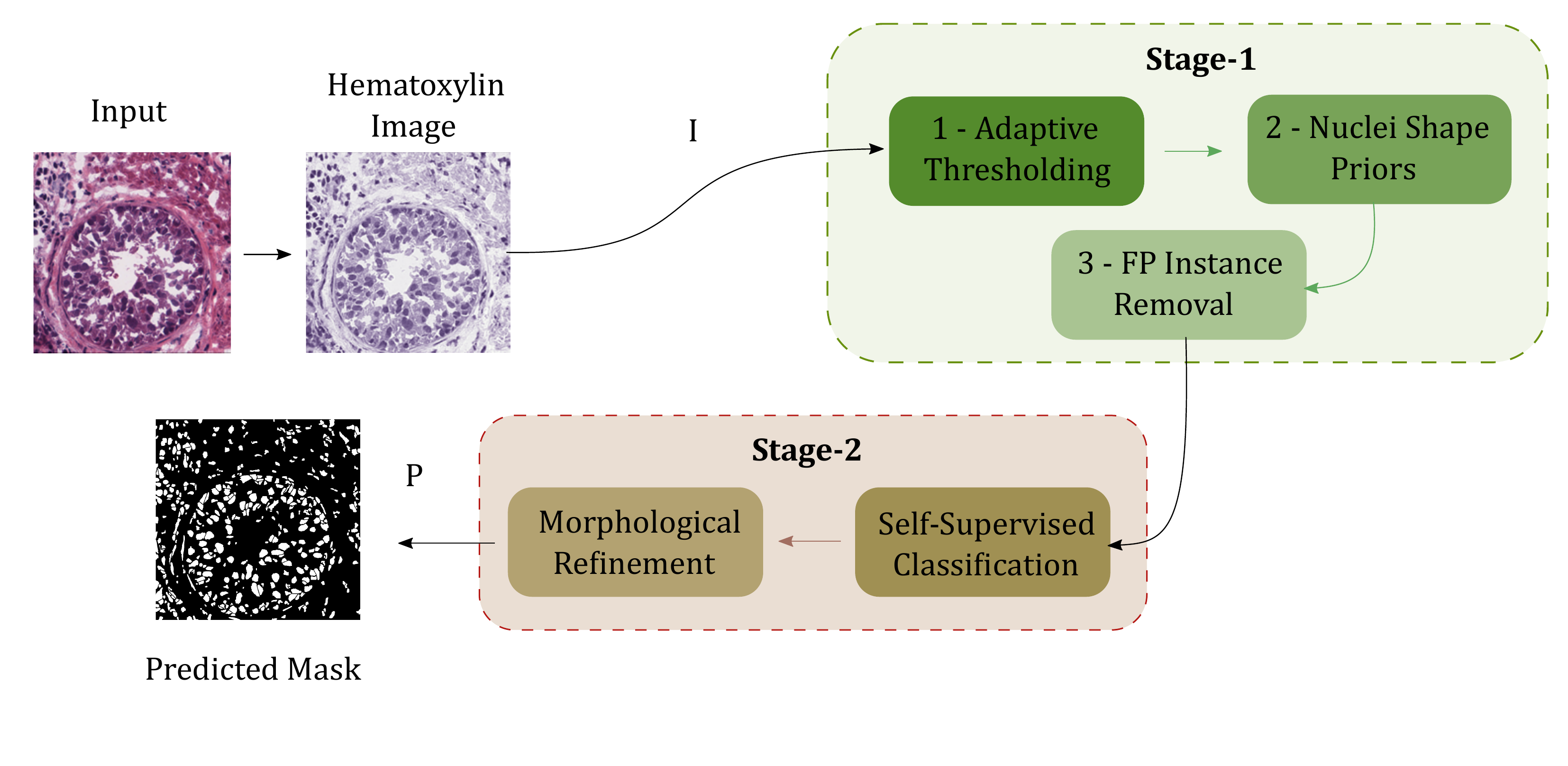}
\end{center}
\vspace{-10mm}
\caption{An overview of the proposed HUNIS method, where the first stage
provides an initial nuclei segmentation result and yields pseudo-labels
to guide the segmentation in the second stage.} \label{fig:Pipeline}
\end{figure*}

\section{Proposed HUNIS Method}\label{sec:PropMeth}

An overview of the proposed HUNIS method is shown in Fig.
\ref{fig:Pipeline}.  It consists of a two-stage block-wise operations pipeline. The
first stage includes: 1) adaptive thresholding of pixel intensity
values, 2) incorporation of nuclei size/shape priors and 3) removal of
false positive instances. The first stage provides an initial
segmentation result and yields the  pseudo-labels for self-supervising the second stage.  The second stage exploits color and shape information under the
self-supervised setting for a more accurate segmentation. 

\subsection{First-Stage Processing}\label{sec:1-1}

\subsubsection{Adaptive Thresholding}\label{sec:1A}

A histology image of size $1000 \times 1000$ is first decomposed into
non-overlapping blocks of size $50\times50$ to ensure homogeneity at the
local level. As a pre-processing step, we accentuate the foreground
nuclei over background tissue, thus enhancing the subsequent adaptive
thresholding operation. In general, the nuclei chromatic palette is
mostly captured from Hematoxylin ($H$), in contrast with Eosin ($E$) that
carries more information about the background. To this end, the original
color image is projected on the $H$ color basis using the approach
described in \cite{dorado2017color}. A contrast enhancement is applied
to the $H$-image to further highlight the nuclei.  Next, color pixels
are converted into monochrome ones within each block to facilitate the
following thresholding operation. The color transformation can be
achieved by applying principal component analysis (PCA) and retaining
the first principal component, called the intensity value below. PCA
removes the correlation among RGB channels and achieves high energy
compaction. The $L$ value in the LAB color space is used as reference to
select the sign of the eigenvector at each block uniquely.  The
color-to-intensity transformation simplifies a color-based segmentation
mechanism from $3$D to $1$D. While it works well for the majority of blocks, we consider color attributes in stage-$2$ to further increase the segmentation performance.

\begin{figure}[t]
\begin{center}
\includegraphics[width=0.7\linewidth]{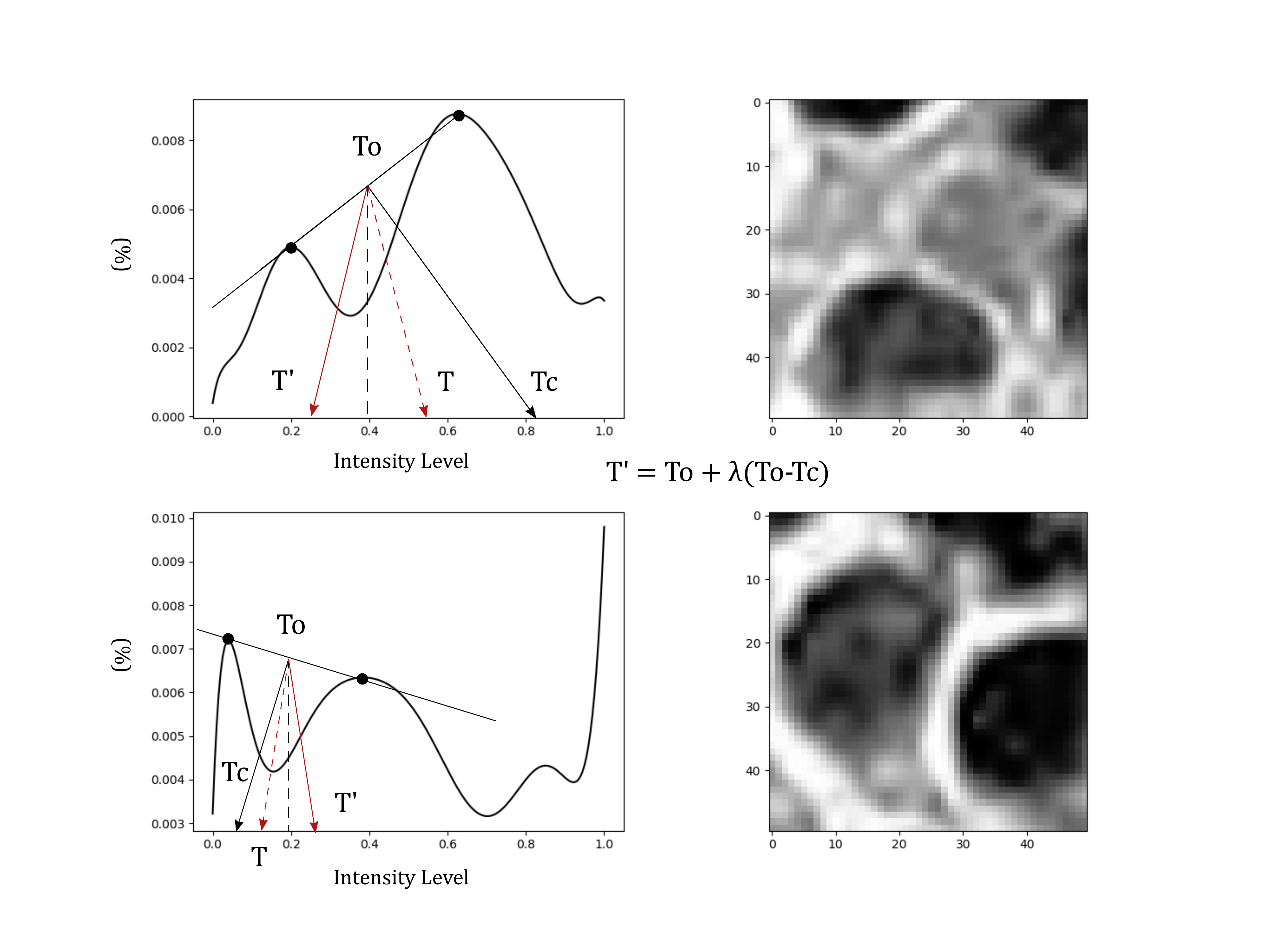}
\end{center}
\vspace{-5mm}
\caption{Two threshold adjustment scenarios: (top) a histogram of two
imbalanced modalities and its corresponding block image and (bottom) a
histogram of three modalities and its corresponding block image.}
\label{fig:Histograms}
\end{figure}

We conduct local thresholding on pixel intensities in each block
adaptively based on a bi-modal assumption. That is, if the intensity
histogram in a block has two main peaks, one corresponding to foreground
and one to background, and the notch between the two peaks is low
enough, one can choose the intermediate point between the two peaks
point as a binarization threshold. There are however challenging cases
where the bi-modal assumption is violated. Then, a mechanism is needed
to adjust the threshold.  Two such examples are shown in Fig.
\ref{fig:Histograms}.  They occur because of ambiguous instances with mid-level intensities (see the top case) or poor contrast between nuclei and background (see the bottom case). 

The threshold is adjusted using the following algorithm.  The peak
points of the first and second modalities are shown by solid dots in the
left two sub-figures of Fig.  \ref{fig:Histograms}. The mid-point between
the two is denoted by $T_o$. We draw a line that is perpendicular to the
line segment formed by the two dots and passing through $T_0$.  Its
intercept, $T_c$, with the horizontal line of zero occurrence is used as
a reference point for correcting the initial $T_o$. Then, we have
\begin{equation}\label{eq:Thres_Cor}
T' = T_o + \lambda(T_o - T_c), \quad 0 < \lambda < 1,
\end{equation}
where $T'$ is the adjusted threshold value and the second term on the
left-hand-side is called a correction term. As shown in the top case of
Fig. \ref{fig:Histograms}, if the second peak (i.e., background) is
higher than the first one (i.e., nuclei), it is likely that the block
has low contrast and ambiguous regions and it is desired to decrease
$T_o$ to reduce the false positive rate.  On the other hand, as shown in
the bottom case of Fig. \ref{fig:Histograms}, if the mid-peak is high
but not higher than the first peak (i.e., nuclei), it is likely that the
mid-modality region corresponds to nuclei boundaries (given the first
peak being the nuclei) or due to its texture. Thus, we can increase
$T_o$ to segment nuclei more precisely. 

The direction and magnitude of threshold adjustment is automatically
determined from the slope of the line segment connecting the two dots. A
positive slope would give an intercept, $T_c$, higher from $T_o$ and the
correction term in Eq. (\ref{eq:Thres_Cor}) is negative.  Conversely, a
negative slope would yield a positive correction.  A weight, $\lambda$,
is used to control the correction amount, whose value is obtained
empirically. Note that point $T$ in Fig. \ref{fig:Histograms} is the
reflection of $T'$ about the vertical line with intensity equal to
$T_o$. It does not appear in Eq. (\ref{eq:Thres_Cor}) and it is drawn only for illustration purposes. 

\subsubsection{Incorporation of Size/Shape Priors}\label{sec:1B}

The output from the adaptive thresholding module is often noisy.  Prior
knowledge about nuclei size and shape can be incorporated to remove that
noisy predictions at the instance level. To this end, we calculate the histogram of nuclei sizes in an
image. Unusually small nuclei instances from unsupervised thresholding are
perceived as noise and can be filtered out. Moreover, two or more close nuclei can
end up being connected after segmentation because of unclear boundaries. Shape
priors (e.g., nuclei shape convexity) can help split those falsely
connected nuclei. Algorithmically, the convex hull algorithm can detect
abnormally steep curves along nucleus boundary, indicating that the
instance came after two or more bundled nuclei and so they can be split.
Furthermore, hole filling is used to correct open areas in the interior
of a nucleus and thus compensate for inner texture variations that
challenge the thresholding operation. 

\subsubsection{Removal of False Positive Instances}\label{sec:1C}

Some false positive nuclei instances cannot be filtered out in the
size/shape priors module. They usually come from darker background areas
(resulting from defects in staining process) or small nuclei with
ambiguous texture. We propose a simple and efficient way to reduce
the false positive rate. In this module, we consider a larger local
neighborhood, called a tile (say, of size $200\times200$) to include more
detected instances for consideration. Each tile is processed
independently. The idea is to compare instances that are more likely to
be true positives with other ambiguous instances that could potentially be false
positives.  The size prior contributes here. Larger instances are
less likely to be false positives while the chances for smaller instances
to be falsely marked are higher. This is mainly due to the combined operation of adaptive threshoding and size/priors modules. Hence, we have a global nuclei size
threshold to deduce what instances will be used as ``ground truth" in a
tile. To be more specific, we have two sets of instances: the reference
instances set, $R$, and the query instances set, $Q$. All instances with
sizes larger than a threshold are assigned to set $R$ while the
remaining ones to set $Q$.  Each element in $Q$ is compared against the mean ensemble of elements in $R$. If their similarity is poor, it is likely to be a FP instance and can be eliminated. 

\begin{figure}[t]
\begin{center}
\includegraphics[width=0.6\linewidth]{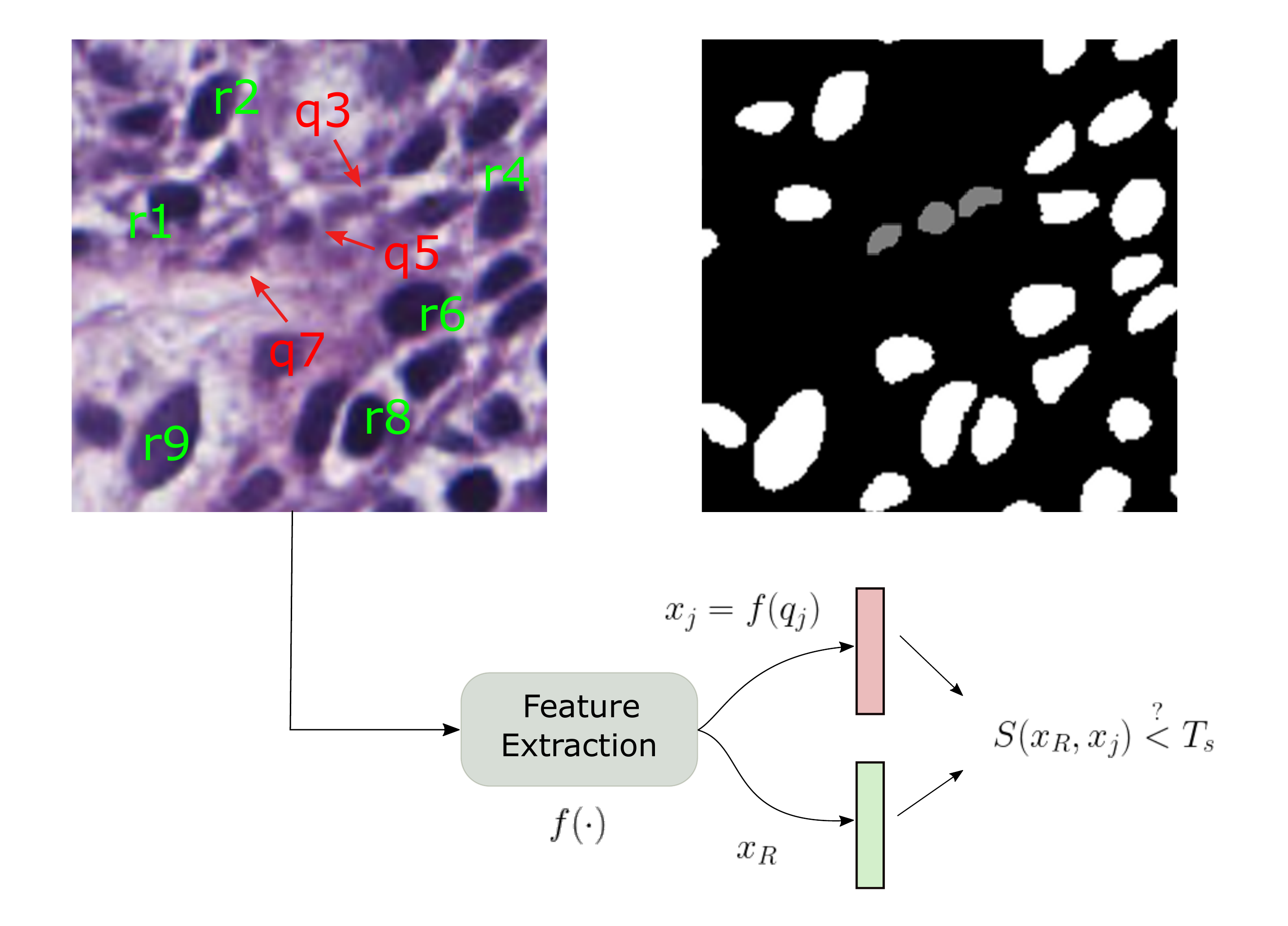}
\end{center}
\vspace{-5mm}
\caption{Illustration of the effect of the false positive removal module,
where some small-sized instances (in red) are compared with larger instances 
(in green) that are more likely to be actual nuclei. The marked instances in grey (right sub-figure) have a similarity score below $T_s$ and are eliminated.} \label{fig:FPReduction}
\end{figure}

To evaluate the similarity and compare instances in $R$ and $Q$, some attributes are extracted per
instance. Since most instances from $Q$ have lower contrast and poorer color saturation, we use HSV colorspace and the corresponding contrast value per instance as features to discern FP nuclei. For $R$ class, we aggregate all instance features to yield one reference feature for comparison.   That is, we extract the feature vector ${\bf
x}_R$ by averaging their values of all instances in $R$ via
\begin{equation}\label{eq:Aggreg}
{\bf x}_R = \frac{1}{|R|} \sum_{i \in R} {\bf x}_i,
\end{equation}
where ${\bf x}_i$ is the feature vector of the $i$-th instance in $R$.
Then, we compare the similarity of the same feature vector of a query 
sample against ${\bf x}_R$. The similarity metric is defined as:
\begin{equation}\label{eq:Kernel}
S({\bf x}_R, {\bf x}_j) = e^{-\gamma||{\bf x}_j - {\bf x}_R||^2_2}, 
\quad \forall j \in Q,
\end{equation}
where $\gamma$ is a hyper-parameter. Clearly, $0 \le S \le 1$. The higher the
$S$ value, the higher the similarity. A query instance is removed, if
its $S < T_S$, where $T_S$ is another hyperparameter.  The process of
false-positive nuclei instance removal is illustrated in Fig.
\ref{fig:FPReduction}. 

\subsection{Second-Stage Processing}\label{sec:1-2}

\subsubsection{Self-Supervised Binary Classification for Uncertain Pixels}\label{sec:2A}

The nuclei color distribution in a tile is more stable than that in the entire image and gives richer information about nuclei appearance over background. As such, unlike stage-$1$ that carries out operations on monochrome patches, this module operates in tiles of $200\times200$ from the Hematoxylin image. We first train a classifier based on the pseudo-labels obtained from step-$1$ in a pixel-wise manner. Then, we conduct prediction for uncertain pixels, based on the classifier's confidence. Most of those pixels usually lie close to the nuclei boundaries. As long as the majority of pixel labels in a tile are correct, some of the noise in pseudo-labels can be removed and thus pixels are more likely to be assigned to their correct class. This implies that large and solid nuclei are more probable to stay intact, while some of the smaller instances or nuclei boundary areas can be corrected towards their correct class.  

\subsubsection{Shape Refinement}\label{sec:2B}

In this module, we perform a final round of nuclei shape refinement
using the same priors and procedures as in its counterpart from Stage-1.
Since pixel-wise classification is prone to false predictions to some
extent, this module refines the output so that it has a better segmented nuclei instance. This refinement is necessary since the shapes of some
nuclei could be distorted after splitting or due to unclear boundaries. Thus, we preserve convexity using the convex hull algorithm for nuclei
that have abnormally steep contours regions. Other morphological
operations are also used to refine the nuclei shape. 

\section{Experimental Results}\label{sec:Results}

The proposed HUNIS method is evaluated on the 2018 MICCAI MoNuSeg
dataset \cite{kumar2017dataset} to demonstrate its effectiveness. The
dataset offers different testing protocols.  For performance
benchmarking, we follow the data splitting scheme as specified in
\cite{kumar2017dataset}.  That is, we report results on two MoNuSeg
Challenge test datasets:
\begin{itemize}
\item 14 images from various organs whose histology images are available 
in the training dataset, here referred to as MoNuSeg Test-1. 
\item 6 histology images from three unseen organs (bladder, colon
and stomach), referred to as MoNuSeg Test-2. 
\end{itemize}
The parameters in Eqs.  (\ref{eq:Thres_Cor}) and (\ref{eq:Kernel}) are
set to $\lambda = 0.3$ and $\gamma=0.1$, respectively. The similarity
threshold in Fig.  \ref{fig:FPReduction} is set to $T_s=0.6$. All other
parameters are determined automatically from the data. For evaluation purposes, it is common among other works to
use the Aggregated Jaccard Index (AJI) \cite{kumar2017dataset}, rather than the F-$1$ score or the DICE coefficient. AJI is more suitable for instance-level segmentation tasks, since it takes both nuclei-level
detection and pixel-level error performance into account. 

\begin{table}[h]
\caption{Quantitative results and performance comparison on 
Test-$1$ using AJI metric.} \label{tab:CompResults_Challg}
\begin{center}
\begin{tabular}{|c c|} \hline
\hspace{2mm} Method \hspace{2mm}  & \hspace{5mm}  AJI  \hspace{5mm}   \\ \hline
\multicolumn{2}{|c|}{{\em Unsupervised}} \\ \hline
\rule[-1ex]{0pt}{3.5ex}  DARCNN \cite{hsu2021darcnn} & 0.4461   \\ \hline
\rule[-1ex]{0pt}{3.5ex}  Hou {\em et al.} \cite{hou2019robust} & 0.4980 \\ \hline
\rule[-1ex]{0pt}{3.5ex}  Self-Supervised \cite{sahasrabudhe2020self} &  0.5354 \\ \hline
\rule[-1ex]{0pt}{3.5ex}  Liu {\em et al.} \cite{liu2020unsupervised} &  0.5610 \\ \hline
\rule[-1ex]{0pt}{3.5ex}  CBM \cite{magoulianitis2021unsupervised} &  0.6142 \\ \hline
\rule[-1ex]{0pt}{3.5ex}  \textbf{HUNIS (ours)}  & \bf{0.6387}   \\ \hline
\multicolumn{2}{|c|}{{\em Supervised}} \\ \hline
\rule[-1ex]{0pt}{3.5ex}  CNN3 \cite{kumar2017dataset}  & 0.5083  \\ \hline
\rule[-1ex]{0pt}{3.5ex}  Hover-Net \cite{graham2019hover} & 0.618 \\ \hline
\rule[-1ex]{0pt}{3.5ex}  UNet-Atten. \cite{schlemper2019attention} & 0.6498  \\ \hline
\rule[-1ex]{0pt}{3.5ex}  NucleiSegNet \cite{lal2021nucleisegnet} & \bf{0.688}   \\ \hline
\end{tabular}
\end{center}
\end{table}

The results for MoNuSeg Test-1 and Test-2 are shown in Tables
\ref{tab:CompResults_Challg} and \ref{tab:CompResults_Test2},
respectively. All benchmarking methods except CBM are DL-based.  As
shown in Table \ref{tab:CompResults_Challg}, one can see that HUNIS
outperforms all DL unsupervised approaches by large margins in Test-1.
It also outperforms CBM by 0.0245 in terms of the AJI score.
Furthermore, HUNIS achieves a competitive standing among
state-of-the-art supervised DL methods in Test-1. Its performance is
close to that of the 2nd best in the table, namely, UNet-Atten.
\cite{schlemper2019attention}. 

The domain adaptation task in Test-2 is quite challenging for supervised
methods since they need to make a decision on data from unseen organ.  As shown in
Table \ref{tab:CompResults_Test2}, HUNIS outperforms all benchmarking
unsupervised and supervised methods, including sophisticated DL models
such as CIA-Net.  Evidently, it is difficult for DL models to generalize
well from training to testing when the amount of annotated data is
scarce.  In contrast, Test-1 and Test-2 make little difference for
unsupervised methods. We see that HUNIS can achieve an even higher AJI
score in Test-2 as compared with Test-1. 

\begin{table}[h]
\caption{Quantitative results and performance comparison on Test-$2$ 
using the AJI metric.}\label{tab:CompResults_Test2}
\begin{center}
\begin{tabular}{|c c|} \hline
\hspace{5mm} Method \hspace{5mm}  & \hspace{5mm} AJI   \hspace{5mm}  \\ \hline
\multicolumn{2}{|c|}{{\em Unsupervised}} \\ \hline
\rule[-1ex]{0pt}{3.5ex}  Cell Profiler \cite{carpenter2006cellprofiler}  & 0.0809  \\ \hline
\rule[-1ex]{0pt}{3.5ex}  Fiji \cite{schindelin2012fiji} & 0.3030   \\ \hline
\rule[-1ex]{0pt}{3.5ex}  CBM \cite{magoulianitis2021unsupervised} &  0.5808 \\ \hline
\rule[-1ex]{0pt}{3.5ex}  \textbf{HUNIS (ours)} & \bf{0.6548}    \\ \hline
\multicolumn{2}{|c|}{{\em Supervised}} \\ \hline
\rule[-1ex]{0pt}{3.5ex}  CNN3 \cite{kumar2017dataset} &  0.4989   \\ \hline
\rule[-1ex]{0pt}{3.5ex}  BES-Net \cite{oda2018besnet} & 0.5823   \\ \hline
\rule[-1ex]{0pt}{3.5ex}  CIA-Net \cite{zhou2019cia} &  \bf{0.6306}  \\ \hline
\end{tabular}
\end{center}
\end{table}

An ablation study that demonstrates the progressive improvement on
segmentation results in various stages of HUNIS is given in Table
\ref{tab:Ablation}.  It shows the effectiveness of false positive
instances removal module (Module 3 in Stage 1) and two modules in Stage
2. They contribute to significant AJI score improvement from the output
of the previous stage. 

It is worthwhile to stress that HUNIS does not carry any learnable
parameters. In contrast, modern DL models typically contain millions of
parameters. For example, the NucleiSegNet model and the UNet-Atten model
take up 15.7M and 32M parameters, respectively.  They need GPU to
conduct the training and testing tasks. HUNIS can be implemented by
software in mobile/edge devices. Furthermore, it is a fully unsupervised
solution requiring no training data at all. 

\begin{table}[h]
\caption{Segmentation improvement over different stages of our pipeline 
in Test-$2$ set}\label{tab:Ablation}
\begin{center}
\begin{tabular}{|c|c|c|c|} \hline
\textbf{Stages}   & Stage-1 (Modules 1\&2) & Stage-1 (Modules 1\&2\&3) &  
Stages 1\&2   \\ \hline
\rule[-1ex]{0pt}{3.5ex}  AJI & 0.6045 & 0.6377 & 0.6548  \\ \hline
\end{tabular}
\end{center}
\end{table}

\section{Conclusion and Future Work}\label{sec:Concls}

An unsupervised nuclei instance segmentation method, namely HUNIS, was
proposed in this work. It contains several novel ideas, such as an
advanced adaptive thresholding scheme that can adjust the binarization
threshold based on the local distribution automatically, an efficient
false positive nuclei removal technique that can eliminate ambiguous
instances, and a self-supervised learning mechanism that can finetune
the segmentation results. Experimental results showed that HUNIS outperforms
other unsupervised methods and maintains competitive performance against
state-of-the-art supervised methods. It has no learnable parameters and
it comes with very low computational complexity, thus offering a green
solution to the nuclei instance segmentation problem. It is interesting
to extend the developed methodology to other relevant medical segmentation
problems as well.

\bibliographystyle{unsrt}  
\bibliography{references}

\begin{thebibliography}{10}

\bibitem{gurcan2009histopathological}
Metin~N Gurcan, Laura~E Boucheron, Ali Can, Anant Madabhushi, Nasir~M Rajpoot,
  and Bulent Yener.
\newblock Histopathological image analysis: A review.
\newblock {\em IEEE reviews in biomedical engineering}, 2:147--171, 2009.

\bibitem{caicedo2019nucleus}
Juan~C Caicedo, Allen Goodman, Kyle~W Karhohs, Beth~A Cimini, Jeanelle
  Ackerman, Marzieh Haghighi, CherKeng Heng, Tim Becker, Minh Doan, Claire
  McQuin, et~al.
\newblock Nucleus segmentation across imaging experiments: the 2018 data
  science bowl.
\newblock {\em Nature methods}, 16(12):1247--1253, 2019.

\bibitem{xue2011t}
Jing-Hao Xue and D~Michael Titterington.
\newblock $ t $-tests, $ f $-tests and otsu's methods for image thresholding.
\newblock {\em IEEE Transactions on Image Processing}, 20(8):2392--2396, 2011.

\bibitem{lu2012robust}
Cheng Lu, Muhammad Mahmood, Naresh Jha, and Mrinal Mandal.
\newblock A robust automatic nuclei segmentation technique for quantitative
  histopathological image analysis.
\newblock {\em Analytical and Quantitative Cytology and Histology},
  34:296--308, 2012.

\bibitem{magoulianitis2021unsupervised}
Vasileios Magoulianitis, Peida Han, Yijing Yang, and C-C~Jay Kuo.
\newblock Unsupervised data-driven nuclei segmentation for histology images.
\newblock {\em arXiv preprint arXiv:2110.07147}, 2021.

\bibitem{hafiane2008fuzzy}
Adel Hafiane, Filiz Bunyak, and Kannappan Palaniappan.
\newblock Fuzzy clustering and active contours for histopathology image
  segmentation and nuclei detection.
\newblock In {\em International Conference on Advanced Concepts for Intelligent
  Vision Systems}, pages 903--914. Springer, 2008.

\bibitem{al2016white}
Khamael Al-Dulaimi, Inmaculada Tomeo-Reyes, Jasmine Banks, and Vinod Chandran.
\newblock White blood cell nuclei segmentation using level set methods and
  geometric active contours.
\newblock In {\em 2016 International Conference on Digital Image Computing:
  Techniques and Applications (DICTA)}, pages 1--7. IEEE, 2016.

\bibitem{roula2004evolutionary}
Mohammed~Ali Roula, Ahmed Bouridane, and Fatih Kurugollu.
\newblock An evolutionary snake algorithm for the segmentation of nuclei in
  histopathological images.
\newblock In {\em 2004 International Conference on Image Processing, 2004.
  ICIP'04.}, volume~1, pages 127--130. IEEE, 2004.

\bibitem{xu2021unsupervised}
Hongming Xu, Lina Liu, Xiujuan Lei, Mrinal Mandal, and Cheng Lu.
\newblock An unsupervised method for histological image segmentation based on
  tissue cluster level graph cut.
\newblock {\em Computerized Medical Imaging and Graphics}, 93:101974, 2021.

\bibitem{shen2015segmenting}
Pengfei Shen, Wenjian Qin, Jie Yang, Wanming Hu, Shifu Chen, Ling Li, TieXiang
  Wen, and Jia Gu.
\newblock Segmenting multiple overlapping nuclei in h\&e stained breast cancer
  histopathology images based on an improved watershed.
\newblock In {\em 2015 IET International Conference on Biomedical Image and
  Signal Processing (ICBISP 2015)}, pages 1--4. IET, 2015.

\bibitem{veta2011marker}
Mitko Veta, A~Huisman, Max~A Viergever, Paul~J van Diest, and Josien~PW Pluim.
\newblock Marker-controlled watershed segmentation of nuclei in h\&e stained
  breast cancer biopsy images.
\newblock In {\em 2011 IEEE international symposium on biomedical imaging: from
  nano to macro}, pages 618--621. IEEE, 2011.

\bibitem{kumar2017dataset}
Neeraj Kumar, Ruchika Verma, Sanuj Sharma, Surabhi Bhargava, Abhishek Vahadane,
  and Amit Sethi.
\newblock A dataset and a technique for generalized nuclear segmentation for
  computational pathology.
\newblock {\em IEEE transactions on medical imaging}, 36(7):1550--1560, 2017.

\bibitem{oda2018besnet}
Hirohisa Oda, Holger~R Roth, Kosuke Chiba, Jure Sokoli{\'c}, Takayuki Kitasaka,
  Masahiro Oda, Akinari Hinoki, Hiroo Uchida, Julia~A Schnabel, and Kensaku
  Mori.
\newblock Besnet: boundary-enhanced segmentation of cells in histopathological
  images.
\newblock In {\em International Conference on Medical Image Computing and
  Computer-Assisted Intervention}, pages 228--236. Springer, 2018.

\bibitem{graham2019hover}
Simon Graham, Quoc~Dang Vu, Shan E~Ahmed Raza, Ayesha Azam, Yee~Wah Tsang,
  Jin~Tae Kwak, and Nasir Rajpoot.
\newblock Hover-net: Simultaneous segmentation and classification of nuclei in
  multi-tissue histology images.
\newblock {\em Medical Image Analysis}, 58:101563, 2019.

\bibitem{lal2021nucleisegnet}
Shyam Lal, Devikalyan Das, Kumar Alabhya, Anirudh Kanfade, Aman Kumar, and
  Jyoti Kini.
\newblock Nucleisegnet: robust deep learning architecture for the nuclei
  segmentation of liver cancer histopathology images.
\newblock {\em Computers in Biology and Medicine}, 128:104075, 2021.

\bibitem{schlemper2019attention}
Jo~Schlemper, Ozan Oktay, Michiel Schaap, Mattias Heinrich, Bernhard Kainz, Ben
  Glocker, and Daniel Rueckert.
\newblock Attention gated networks: Learning to leverage salient regions in
  medical images.
\newblock {\em Medical image analysis}, 53:197--207, 2019.

\bibitem{zhou2019cia}
Yanning Zhou, Omer~Fahri Onder, Qi~Dou, Efstratios Tsougenis, Hao Chen, and
  Pheng-Ann Heng.
\newblock Cia-net: Robust nuclei instance segmentation with contour-aware
  information aggregation.
\newblock In {\em International Conference on Information Processing in Medical
  Imaging}, pages 682--693. Springer, 2019.

\bibitem{xie2020instance}
Xinpeng Xie, Jiawei Chen, Yuexiang Li, Linlin Shen, Kai Ma, and Yefeng Zheng.
\newblock Instance-aware self-supervised learning for nuclei segmentation.
\newblock In {\em International Conference on Medical Image Computing and
  Computer-Assisted Intervention}, pages 341--350. Springer, 2020.

\bibitem{sahasrabudhe2020self}
Mihir Sahasrabudhe, Stergios Christodoulidis, Roberto Salgado, Stefan Michiels,
  Sherene Loi, Fabrice Andr{\'e}, Nikos Paragios, and Maria Vakalopoulou.
\newblock Self-supervised nuclei segmentation in histopathological images using
  attention.
\newblock In {\em International Conference on Medical Image Computing and
  Computer-Assisted Intervention}, pages 393--402. Springer, 2020.

\bibitem{liu2020unsupervised}
Dongnan Liu, Donghao Zhang, Yang Song, Fan Zhang, Lauren O'Donnell, Heng Huang,
  Mei Chen, and Weidong Cai.
\newblock Unsupervised instance segmentation in microscopy images via panoptic
  domain adaptation and task re-weighting.
\newblock In {\em Proceedings of the IEEE/CVF conference on computer vision and
  pattern recognition}, pages 4243--4252, 2020.

\bibitem{hsu2021darcnn}
Joy Hsu, Wah Chiu, and Serena Yeung.
\newblock Darcnn: Domain adaptive region-based convolutional neural network for
  unsupervised instance segmentation in biomedical images.
\newblock In {\em Proceedings of the IEEE/CVF Conference on Computer Vision and
  Pattern Recognition}, pages 1003--1012, 2021.

\bibitem{hou2019robust}
Le~Hou, Ayush Agarwal, Dimitris Samaras, Tahsin~M Kurc, Rajarsi~R Gupta, and
  Joel~H Saltz.
\newblock Robust histopathology image analysis: To label or to synthesize?
\newblock In {\em Proceedings of the IEEE/CVF Conference on Computer Vision and
  Pattern Recognition}, pages 8533--8542, 2019.

\bibitem{dorado2017color}
Paula~Andrea Dorado, Raul Celis, and Eduardo Romero.
\newblock Color separation of h\&e stained samples by linearly projecting the
  rgb representation onto a custom discriminant surface.
\newblock In {\em 12th International Symposium on Medical Information
  Processing and Analysis}, volume 10160, page 101600P. International Society
  for Optics and Photonics, 2017.

\bibitem{carpenter2006cellprofiler}
Anne~E Carpenter, Thouis~R Jones, Michael~R Lamprecht, Colin Clarke, In~Han
  Kang, Ola Friman, David~A Guertin, Joo~Han Chang, Robert~A Lindquist, Jason
  Moffat, et~al.
\newblock Cellprofiler: image analysis software for identifying and quantifying
  cell phenotypes.
\newblock {\em Genome biology}, 7(10):1--11, 2006.

\bibitem{schindelin2012fiji}
Johannes Schindelin, Ignacio Arganda-Carreras, Erwin Frise, Verena Kaynig, Mark
  Longair, Tobias Pietzsch, Stephan Preibisch, Curtis Rueden, Stephan Saalfeld,
  Benjamin Schmid, et~al.
\newblock Fiji: an open-source platform for biological-image analysis.
\newblock {\em Nature methods}, 9(7):676--682, 2012.

\end{thebibliography}

\end{document}